\documentclass[12pt]{article}
\usepackage[dvips]{graphicx}
\usepackage{amsfonts}
\usepackage{amsmath}
\usepackage{textcomp}
\usepackage{tabularx}
\usepackage{moreverb}
\usepackage{epsfig}
{\catcode`\|=\active \gdef|{\egroup\,\vrule\,\bgroup}}

\newcommand{\Tr}{\textrm{Tr }}

\newcolumntype{Y}{>{\raggedleft\arraybackslash}X}

\begin{document}
\title{Extensive ground state entropy in supersymmetric lattice models}
\author{Hendrik van Eerten$^1$ 
\medskip \\
$^1$Instituut voor Theoretische Fysica, Universiteit van Amsterdam, \\ Valckenierstraat 65, 1018 XE Amsterdam, The Netherlands\\
{\tt hveerten@science.uva.nl}}
\maketitle
\begin{abstract}
We present the result of calculations of the Witten index for a supersymmetric lattice model on lattices of various type and size. Because the model remains supersymmetric at finite lattice size, the Witten index can be calculated using row-to-row transfer matrices and the calculations are similar to calculations of the partition function at negative activity -1. The Witten index provides a lower bound on the number of ground states. We find strong numerical evidence that the Witten index grows exponentially with the number of sites of the lattice, implying that the model has extensive entropy in the ground state.
\end{abstract}
\newpage
\section{Introduction}
In \cite{Schoutens} a spinless fermion lattice model was proposed that is $N=2$  supersymmetric (SUSY) regardless of lattice size or shape. The (fine-tuned) Hamiltonian of this model is written in terms of fermion creation and annihilation operators, as follows:
\begin{equation}
H = \sum_i \sum_{j \textrm{ next to } i} P_{<i>} c_i^\dagger c_j P_{<j>} +\sum_i P_{<i>},
\label{Hamiltonian}
\end{equation}
with $i$ numbering the lattice sites, $c_i$ and $c_i^\dagger$ obeying $\lbrace c_i, c_j^\dagger \rbrace = \delta_{ij}$ and where we have introduced \emph{projection operators P} for convenient notation. The latter have the form:
\begin{equation}
P_{<i>} \equiv \prod_{j \textrm{ next to } i} (1 - c_j^\dagger c_j ).
\end{equation}
These projection operators ensure that \emph{no two neighbouring sites can be occupied simultaneously}. Therefore, we will work with a restricted Hilbert space where these states are excluded.

The SUSY model is closely related to more conventional lattice models, like the Heisenberg XXZ chain. This relation, which holds if we take a 1D chain with special boundary conditions for the lattice, was demonstrated in \cite{FNSLat}. There it was also shown how SUSY could aid a Bethe Ansatz computation of the spectrum of the XXZ chain. Indeed, it is fruitful to see how SUSY can facilitate calculations on the lattice model specified by (\ref{Hamiltonian}), especially since the symmetry holds on arbitrary graphs while exact results on higher dimensional lattices are notoriously difficult to obtain using conventional methods.

The supersymmetry can be viewed as a multiple site generalization of supersymmetric quantum mechanics. The two generators $Q^+$ and $Q^-$ are given explicitly in \cite{Schoutens}:
\begin{equation}
Q^+ = \sum_{i=1}^{N} c_i^\dagger P_{<i>}, \qquad
Q^- = \sum_{i=1}^N c_i P_{<i>}.
\end{equation}
We have the following algebraic structure (with $F$ the fermionic number operator):
\begin{equation}
[F, Q^\pm ] = \pm Q^{\pm}, \quad  [F, H] = 0, \quad
\lbrack Q^+, H \rbrack = 0, \quad \lbrack Q^-, H \rbrack = 0, \quad 
\lbrace Q^+, Q^- \rbrace = H.
\end{equation}

\begin{figure}
\begin{center}
\includegraphics[width=3 cm, height=3cm]{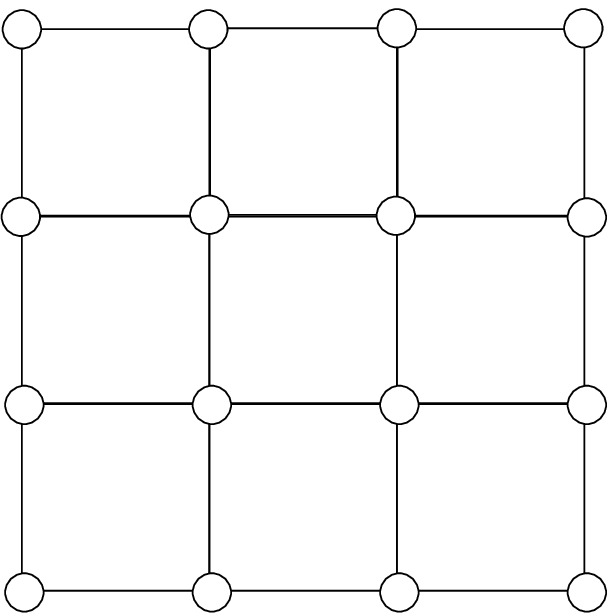}
\includegraphics[width=3 cm, height=3cm]{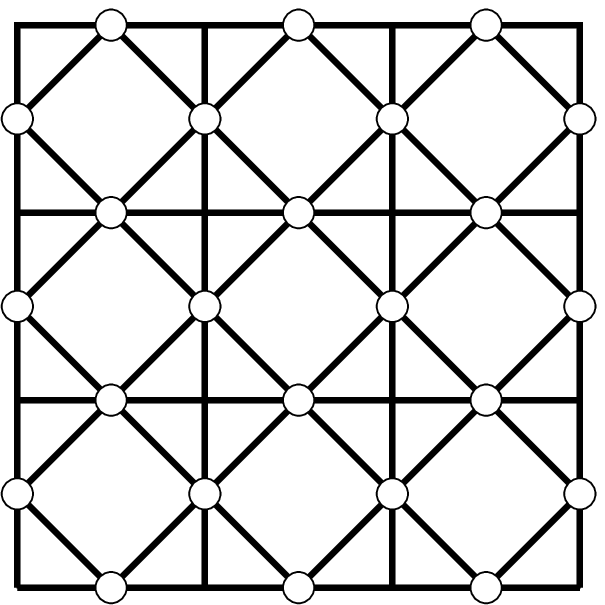}
\caption{The square lattice and the square dimer lattice}
\label{square_lattices}

\includegraphics[width=3 cm, height=3cm]{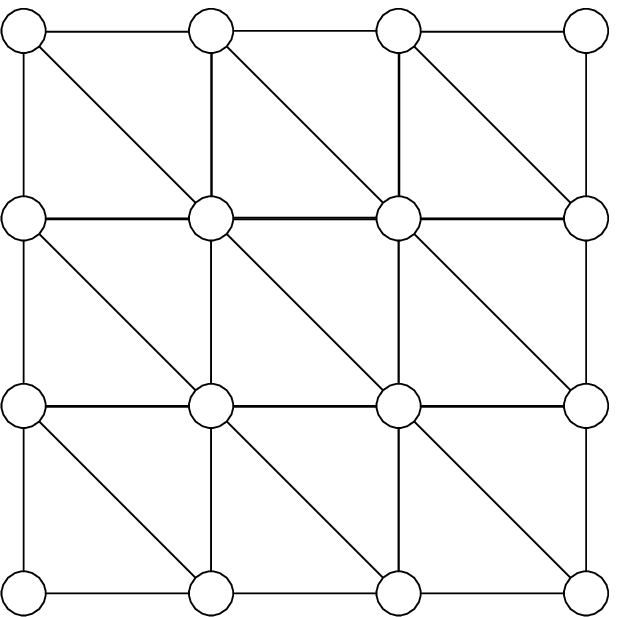}
\includegraphics[width=3 cm, height=3cm]{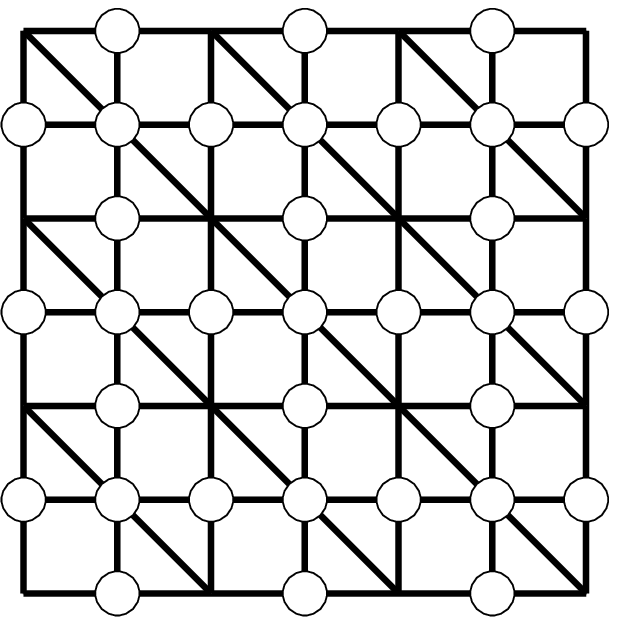}
\caption{The triangular lattice and the triangular dimer lattice}

\includegraphics[width=3 cm, height=3cm]{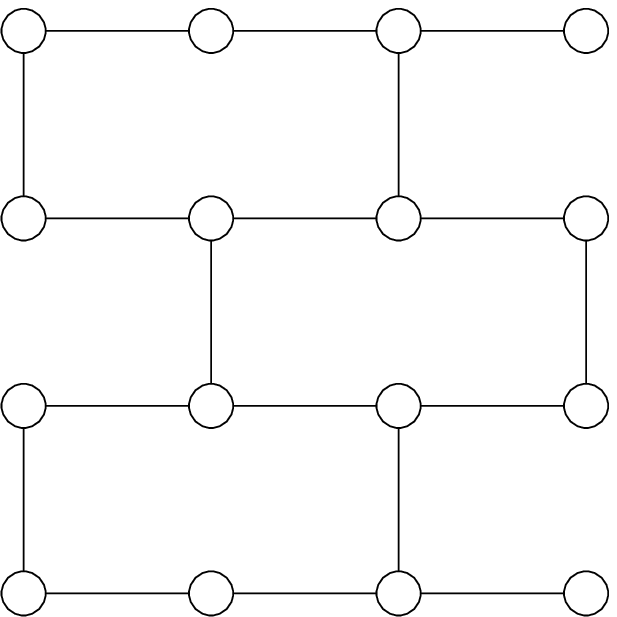}
\includegraphics[width=3 cm, height=3cm]{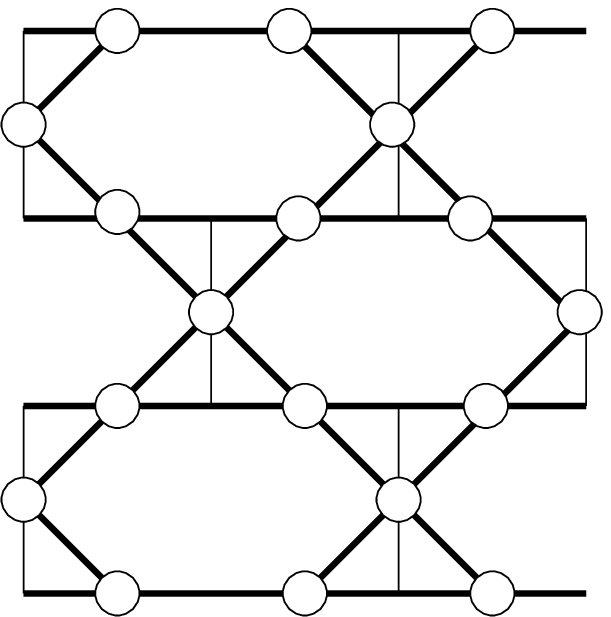}
\caption{The hexagonal lattice and the hexagonal dimer lattice}
\label{hexagonal_lattices}
\end{center}
\end{figure}

In this article we will focus on the ground states of a variety of 2D lattices. We will treat the triangular and hexagonal lattice and their counterparts with dimer configurations as well as the dimer version of the square lattice. The square lattice was treated in a separate publication \cite{Eerten}. All the lattices treated here have an interesting common feature: the number of groundstates of the SUSY Hamiltonian on these lattices increases exponentially with increasing lattice size. This directly implies an extensive entropy of the ground state, a rare property exhibited by only a few condensed matter systems. The lattices are shown in figures \ref{square_lattices} to \ref{hexagonal_lattices}. 

All excited states in the SUSY model occur pairwise, because for every bosonic configuration we have an accompanying fermionic configuration. Here, lattice configurations with an even (odd) number of fermions are termed bosonic (fermionic) configurations. The ground states need not occur pairwise but are restricted otherwise, for their energies are all zero. All this means that we can use the Witten index operator to obtain a lower limit on the number of groundstates:
\begin{equation}
W = \Tr (-1)^F e^{- \beta H}.
\end{equation}
The pairwise cancelling of all excited states allows us to look just at the limiting case $\beta \to 0$, especially when calculating $W$ for finite size lattices where convergence is not an issue, and we will do so for the remainder of this article.

The operator $W$ is completely equivalent to the well known grand canonical partion function $Z = \Tr z^F \exp ( - \beta H)$ of the classical hard particle model in the same dimension, only with a \emph{negative} value for the activity $z$. It is this equivalence that puts a calculation of the Witten index for SUSY lattice models in a broader context and allows us to use well-known approaches for the calculation of $Z$ -especially the use of \emph{transfer matrices} that allows us to write $W$ as the trace of a matrix product. 

We have calculated the Witten index for finite sized lattices up to $15 \times 15$ (the triangular lattice) using transfer matrices. Before discussing the resulting values for the Witten indices, we will first explain how these matrices were constructed.

\section{The transfer matrices}

Transfer matrices can be constructed in various ways, as long as they provide a means of systematically summing over all possible configurations on the lattice (see \cite{Baxter} for an exhaustive introduction). For the $N$-site chain with periodic boundary conditions, the Witten index operator can be written in terms of transfer matrices as
\begin{equation}
W = \Tr \left( \begin{array}{cc} 1 & i \\ i & 0 \end{array} \right)^N = \Tr \left( \begin{array}{cc} 1 & -1 \\ 1 & 0 \end{array} \right)^N.
\end{equation}
The upper left entry of the matrix corresponds to adding an empty site to the chain with an empty site at the end, the upper right entry to adding a filled site to an empty site, the lower left entry to adding an empty site to a filled site and the lower right entry to adding a filled site to a filled site. This last entry has to be zero because no configurations with neighbouring sites filled were allowed. To avoid an overcounting per fermion we can either use $-1, 1$ or $i, i$ for the off-diagonal entries, as long as their product is equal to -1. The resulting transfer matrices can be made symmetric, but never Hermitean.

The procedure can easily be generalized to higher dimensional lattices. The easiest strategy is to use row-to-row transfer matrices and have each matrix entry correspond to a specific row configuration. Because the size of these matrices increases fast with increasing lattice size, the actual constructions of the matrices are best left to a computer (which we did).   In the following, we will briefly explain how the matrices were constructed for the various lattices.

The square lattice is the simplest. For each matrix entry $(i,j)$ the computer program just compares the configuration denoted by $i$ with the one denoted by $j$. The matrix entry will be zero if the two configurations have one or more neighbouring vertices occupied when viewed next to each other. If the combination of $i$ and $j$ is allowed, a factor $(-1)^{f(j)}$ is written down, where $f(j)$ stands for the number of fermions in configuration $j$. This means we obtain a non-symmetric matrix with real numbers only. Using a mapping from the different configurations to binary numbers simplifies matters even further (e.g. $\bullet \circ \bullet \circ \circ \to 10100$ etc.). The comparison between the $i$-th and $j$-th configuration can now be performed using a binary \verb+AND+ instruction.

The program is easily modified for the triangular lattice. There are two extra edges per vertex connecting to vertices on a higher or lower row. They are similar to the vertical edge, but with all vertices of the lower row shifted one position to the right. In terms of the computer code, when comparing the $i$-th and $j$-th configuration, we apply a binary right shift on the $j$-th binary representation before using \verb+AND+. The program now performs two tests instead of one to ascertain if it can enter a nonzero value.

For the hexagonal lattice we need to use two transfer matrices instead of one. Viewing the hexagonal lattice as a horizontal brick wall, we see that we have two different rows. If we work in units of $2 \times 2$ sites, we can take the transfer matrix that adds \emph{two} rows at once as the matrix product of the two different matrices of the single rows.

The dimer lattices offer another complication. The square dimer lattice has both horizontal and vertical dimers. When constructing the transfer matrix we will sum over the horizontal dimer contributions. We can do this without problems because the horizontal dimers make no contact with the next row. For the triangular dimer lattice, we calculate the transfer matrices in a similar way.

\begin{figure}
\begin{center}
\includegraphics[height=2cm]{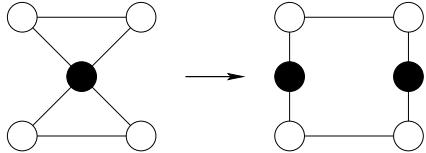}
\caption{Doubling sites on the even rows of the hexagonal dimer lattice}
\label{doubling}
\end{center}
\end{figure}

For the hexagonal dimer lattice we will use yet another trick. As can be seen in figure (\ref{hexagonal_lattices}), the even rows contain only half the number of vertices (after switching dimers and vertices in the original hexagonal lattice). As a first step we shall double this number of vertices in the even rows, like shown in figure (\ref{doubling}). We will have to make certain that the resulting pairs consistently have the same value. Then we can formulate the transfer matrix once more as the product of two transfer matrices.

\section{Results}

The resulting values for the Witten index on various lattice sizes are summarized in the tables in the appendix.

The Witten index for the $M \times N$ lattice is given by $\Tr (\mathbf{T}_N)^M$. Because we are taking the trace, the transfer matrices can be diagonalized without changing the resulting value for the Witten index. As a consequence, in the limiting case where the number of rows goes to infinity we can restrict ourselves to the largest roots of the characteristic polynomial of $\mathbf{T}_N$ (we would have used the term \emph{eigenvalue} if the transfer matrix was not defective). Because the characteristic roots turn out to come in complex pairs, we  have to look at \emph{two} largest roots and we get the following approximation for the Witten index:
\begin{equation}
W_{M,N} \approx a (t_{1,N})^M + \bar{a} (\bar{t}_{1,N})^M,
\end{equation}
where $t_{1,N}$ and $\bar{t}_{1,N}$ denote the complex pair that form the largest roots of the $N$-column transfer matrix.

We can define a \emph{contribution per site} $w$ by
\begin{equation}
w \equiv \lim_{N \to \infty}  \frac{t_{1,N}}{t_{1,N-1}}.
\end{equation}
If we do not take the limit to infinite $N$ but study finite size lattices of increasing size, we find strong numerical evidence that the contribution to the Witten index per site rapidly tends to a fixed complex number that is different for each lattice that was considered. These numbers are shown in table (\ref{indexpersite}). When calculating these numbers we had to select the correct phase factors manually from the two possibilities offered by the complex pairs. The quickly appearing convergence to a fixed value then justified our choice.
\begin{table}[h]
\begin{center}
\begin{small}
\begin{tabular}{l|l|l}
 lattice type       & modulus $| w |$ per site & phase $\arg w$ per site\\ 
\hline
square dimer & $1.33 \pm 0.01$ & $(0.363 \times \pi) \pm 0.01$ \\
triangular & $1.14 \pm 0.01$ & $(0.178 \times \pi) \pm 0.01 $ \\
triangular dimer & $1.261 \pm 0.01$ & $(0.189 \times \pi) \pm 0.001$ \\
hexagonal & $1.2 \pm 0.1$ & insufficient data \\
hexagonal dimer & $1.4 \pm 0.1$ & $(0.310 \times \pi) \pm 0.001$ \\
\end{tabular}
\end{small}
\caption{Witten index per site, various lattice types}
\label{indexpersite}
\end{center}
\end{table}
The square lattice is treated elsewhere \cite{Eerten}, but has the special property that all characteristic roots have norm equal to one. All other lattice types have a single complex pair of largest roots.

This convergence to a complex number greater than 1 has an important consequence, for it indicates an \emph{extensive ground state entropy}. The Witten index can by approximated by
\begin{equation}
W \approx w^{MN} + \bar{w}^{MN} = 2 r^{MN} \cos (MN\theta + \theta_0),
\end{equation}
with $ w = r \exp[ i \theta ]$. The index provides a lower bound on the number of ground states\footnote{Indeed it is very well possible that the actual number of ground states greatly exceeds $| W |$. The reasoning presented here offers us no information on this. See \cite{Eerten}, \cite{Fendley} for an example (the square lattice).}. In our understanding the oscillating (cosine) factor is due to partial cancelations in $W$ of contributions coming from bosonic and fermionic groundstates. We thus expect that the monotonically growing factor $r^{MN}$ sets a lower bound for the actual ground state entropy $S_0 \equiv \ln [ \# \rm{ground} \rm{states}]$ according to
\begin{equation}
S_0 \ge MN \ln r.
\end{equation}
For example, for the triangular lattice we find
\begin{equation}
S_0 \ge MN \ln 1.14 = 0.131 MN.
\end{equation}

It is interesting to see whether the Witten index per site can be calculated analytically.  The triangular lattice case, for example, is equivalent to the hard hexagon model (see \cite{BaxterExact}) with activity -1 and even fits the Yang-Baxter equation at this value. Unfortunately one can not simply modify the reasoning from \cite{Baxter} for the $z < 0$ regime because the transfer matrices are defective, so this question remains open for the moment. It might still be possible to obtain analytical solutions using other methods. In \cite{Fendley}, for example,  an exact solution for the nonagon-triangle lattice was found by showing that it is equivalent to the number of closed packed dimer coverings on the honeycomb lattice.

The fact that we have a complex number indicates that the Witten index does not necessarily grow with increasing lattice size, like the partition sum of the corresponding hard particle model in the $z > 0$ regime would. The phase factor can be used to obtain a rough first estimate on the fermion density of the ground states. The phase factor in the cosine term of $W$ shows us how many sites we need to add to get from a majority of fermionic ground states to a majority of bosonic ground states and vice versa. (for example, the 1D chain has roots $\exp (\pm i/3 \pi)$ and only two ground states both with filling factor $1/3$). But it is only useful as such when combined with a more advanced approach to calculating the spectrum like the spectral sequence technique used in \cite{Fendley}. After all, the phase factor is determined by \emph{all} ground states and they do not necessarily all have the same filling factors.

\section{Acknowledgements}

The author wishes to thank Paul Fendley for helpful discussions and Kareljan Schoutens for careful reading of the manuscript.

\appendix

\section{Witten Index Tables}

The hexagonal lattice tables correspond to the vertical brick wall. It has been confirmed that the tables for the horizontal brick wall are equal to the transposed tables.

\begin{table}[h!]
\begin{footnotesize}
\begin{tabular}{r|rrrrr}
 & 1 & 2 & 3 & 4 & 5 \\
\hline
1 & 1 & -1 & -2 & -1 & 1 \\
2 & -1 & 1 & 8 & -15 & 19 \\
3 & -2 & 8 & -26 & 44 & -92 \\
4 & -1 & -15 & 44 & 129 & -361 \\
5 & 1 & 19 & -92 & -361 & -2344 \\ 
6 & 2 & 4 & 188 & -912 & -9158 \\
7 & 1 & -57 & -338 & 4479 & -24219 \\ 
8 & -1 & 129 & 572 & 4417 & -8241 \\
9 & -2 & -136 & -818 & -46612 & 362068 \\
10 & -1 & -39 & 668 & 5665 & 2617744 \\
\end{tabular}
\end{footnotesize}
\caption{Witten Index for $M \times N$ square dimer lattice}
\end{table}

\begin{table}[h!]
\begin{footnotesize}
\begin{tabular}{r|rrrrr}
 & 6 & 7 & 8 & 9 & 10 \\
 \hline
 1 & 2 & 1 & -1 & -2 & -1 \\
 2 & 4 & -57 & 129 & -136 & -39 \\
 3 & 188 & -338 & 572 & -818 & 668 \\
 4 & -912 & 4479 & 4417 & -46612 & 5665 \\
 5 & -9158 & -24219 & -8241 & 362068 & 2617744 \\
 6 & -54584 & -239790 & -630384 & 1243052 & 31152804 \\
 7 & -239790 & 1495453 & 9803807 & -95946944 & -363241257 \\
 8 & -630384 & 9803807 & -130406911 & 1458639932 & -1665351583 \\
 9 & 1243052 & -95946944 & 1458639932 & -980392698 & -436754324 \\
10 & 31152804 & -363241257 & -1665351583 & -436754324 & 1741554048 \\
\end{tabular}
\end{footnotesize}
\caption{Witten Index for $M \times N$ square dimer lattice (continued)}
\end{table}

\begin{table}[h!]
\begin{footnotesize}
\begin{tabular}{r|rrrrrrrrrrr}
   & 1 &   2 &   3 &     4 &      5 &      6 &       7 &       8 &         9 &        10 \\
\hline
 1 & 1 &   1 &   1 &     1 &      1 &      1 &       1 &       1 &         1 &         1 \\
 2 & 1 &  -3 &  -5 &     1 &     11 &      9 &     -13 &     -31 &        -5 &        57 \\
 3 & 1 &  -5 &  -2 &     7 &      1 &    -14 &       1 &      31 &        -2 &       -65 \\
 4 & 1 &   1 &   7 &   -23 &     11 &     25 &     -69 &     193 &       -29 &      -279 \\
 5 & 1 &  11 &   1 &    11 &     36 &    -49 &     211 &    -349 &       811 &     -1064 \\
 6 & 1 &   9 & -14 &    25 &    -49 &   -102 &     -13 &    -415 &      1462 &     -4911 \\
 7 & 1 & -13 &   1 &   -69 &    211 &    -13 &    -797 &    3403 &     -7055 &      5237 \\
 8 & 1 & -31 &  31 &   193 &   -349 &   -415 &    3403 &     881 &    -28517 &     50849 \\
 9 & 1 &  -5 &  -2 &   -29 &    881 &   1462 &   -7055 &  -28517 &     31399 &    313315 \\
10 & 1 &  57 & -65 &  -279 &  -1064 &  -4911 &    5237 &   50849 &    313315 &    950592 \\
11 & 1 &  67 &   1 &   859 &   1651 &  12607 &   32418 &  159083 &    499060 &   2011307 \\
12 & 1 & -47 & 130 & -1295 &   -589 & -26006 & -152697 & -535895 &  -2573258 &  -3973827 \\
13 & 1 &-181 &   1 &   -77 &  -1949 &  67523 &  330331 & -595373 & -10989458 &  -49705161 \\
14 & 1 & -87 &-257 &  3641 &  12611 &-139935 & -235717 & 5651377 &   4765189 & -232675057 \\
15 & 1 & 275 &  -2 & -8053 & -32664 & 272486 & -1184714&-1867189 & 134858383 & -702709340 \\
\end{tabular}
\end{footnotesize}
\caption{Witten Index for $M \times N$ triangular lattice}
\end{table}

\begin{table}[h!]
\begin{footnotesize}
\begin{tabular}{r|rrrrr}
   &         11 &         12 &          13 &         14 &          15 \\
 \hline
 1 &          1 &          1 &           1 &          1 &           1 \\
 2 &         67 &        -47 &        -181 &        -87 &         275 \\
 3 &          1 &        130 &           1 &       -257 &          -2 \\
 4 &        859 &      -1295 &         -77 &       3641 &       -8053 \\
 5 &       1651 &       -589 &       -1949 &      12611 &      -32664 \\
 6 &      12607 &     -26006 &       67523 &    -139935 &      272486 \\
 7 &      32418 &    -152697 &      330331 &    -235717 &    -1184714 \\
 8 &     159083 &    -535895 &     -595373 &    5651377 &    -1867189 \\
 9 &     499060 &   -2573258 &   -10989458 &    4765189 &   134858383 \\
10 &    2011307 &   -3973827 &   -49705161 & -232675057 &  -702709340 \\
11 &    5102879 &   12409123 &    18205045 & -129877296 & -1457956169 \\
12 &   12409123 &  232286890 &  1851105439 & 1476815313 & -1132095426 \\
13 &   18205045 & 1851105439 & -1938183221 & 1466459831 &  1016873233 \\
14 & -129877296 & 1476815313 &  1466459831 &  139861123 & -1366302204 \\
15 & -1457956169& -1132095426&  1016873233 &-1366302204 &  1417898645 \\
\end{tabular}
\end{footnotesize}
\caption{Witten Index for $M \times N$ triangular lattice (continued)}
\end{table}

\begin{table}[h!]
\begin{footnotesize}
\begin{tabular}{r|rrrrrrrr}
 & 2 & 4 & 6 & 8 & 10 & 12 & 16 & 18 \\
\hline
2  & 1 & -3 & -5 & 1 & 11 & 9 & -13 & -31 \\
4  & -3 & 1 & 21 & -79 & 157 & -71 & -731 & 3105 \\
6  & -5 & 21 & -44 & -47 & 995 & 6576 & 32279 & -131167 \\
8  & 1 & -79 & -47 & 3329 & 3801 & -134959 & -217671 & 5439681 \\
10 & 11 & 157 & 995 & 3801 & -37009 & -1110731 & -17397663 & -217844591 \\ 
12 & 9 & -71 & -6576 & -134959 & -1110731 & 11324392 & 538444825 & 105699937 \\
14 & -13 & -731 & 32279 & -217671 & -17397663 & 538444825 & -600643992 & -914519359\\ 16 & -31 & 3105 & -131167 & 5439681 & -217844591 & 105699937 & -914519359 \\
\end{tabular} \\
\end{footnotesize}
\caption{Witten Index for $M \times N$ triangular dimer lattice}
\end{table}

\begin{table}[h!]
\begin{footnotesize}
\begin{tabular}{r|rrrrrrrrr}
   &  2 &  4 &    6 &     8 &     10 &    12 &      14 &       16 &        18 \\
\hline
 2 & -1 & -1 &    2 &    -1 &    -1 &      2 &      -1 &       -1 &         2 \\
 4 &  3 &  7 &   18 &    47 &   123 &    322 &     843 &     2207 &      5778 \\
 6 & -1 & -1 &   32 &   -73 &    44 &    356 &   -1387 &     2087 &      2435 \\
 8 &  3 &  7 &   18 &    55 &   123 &    322 &     843 &     2215 &      5778 \\
 10 & -1 & -1 &  152 &  -321 &  -171 &   7412 &  -26496 &    10079 &    393767 \\
 12 &  3 &  7 &  156 &  1511 &  6648 &  29224 &  150069 &  1039991 &   6208815 \\
 14 & -1 & -1 &  338 &   727 & -5671 &   1850 &  183560 &  -279497 &  -4542907 \\
 16 &  3 &  7 & 1362 & 12183 & 31803 & 379810 & 5970107 & 55449303 & 327070578 \\
\end{tabular}
\end{footnotesize}
\caption{Witten Index for $M \times N$ hexagonal lattice}
\end{table}

\begin{table}[h!]
\begin{footnotesize}
\begin{tabular}{r|rrrrrrrr}
  &   2 &      4 &        6 &          8 &         10 &          12 &        14 &   16 \\
\hline
2 &   0 &     -8 &        0 &         32 &          0 &       -128 &          0 & 512 \\
4 &  -4 &      8 &       32 &       -224 &        896 &      -2176 &       1536 & 16896 \\
6 &  12 &     88 &      576 &       3296 &      17472 &      77056 &     194304 & -1139200 \\
8 &   4 &   -496 &    -3056 &     118912 &    1287744 &  -25732864 & -439656192 &  626526208 \\
10 & -40 &   1832 &   -42400 &    1088352 &  -19939840 &  205139072 & -878495232 &  1612654080 \\
12 &  44 &  -2872 &  -425344 &  -23115488 &   84888704 &  420235264 &  335598080 & -1677852160 \\
14 &  84 & -12440 & -3459792 & -336941664 & -936816704 & 1524979328 & 1080971264 &  1869085184 \\
\end{tabular}
\end{footnotesize}
\caption{Witten Index for $M \times N$ hexagonal dimer lattice}
\end{table}

\end{document}